\documentclass{aa}  

\usepackage{graphicx, wrapfig, lipsum}
\usepackage{caption}
\usepackage{subcaption}
\usepackage{makecell}
\usepackage{url}
\usepackage{multirow}
\usepackage[export]{adjustbox}
\usepackage[rightcaption]{sidecap}
\usepackage{textcmds}
\usepackage{footnote}
\usepackage{nicematrix}
\usepackage{tikz}
\usetikzlibrary{fit}
\makesavenoteenv{tabular}
\usepackage{placeins}
\usepackage{txfonts}
\usepackage{natbib}
\bibpunct{(}{)}{;}{a}{}{,}
\usepackage[colorlinks=true,linkcolor=blue, citecolor=blue, urlcolor=blue]{hyperref}

\begin{document} 

   \title{Comprehensive study of solar type~II radio bursts and the \\ properties of the associated shock waves}

   \author{K. Bhandari
          \inst{1}, 
          D. E. Morosan \inst{1,2}, S. Normo \inst{1}
          }

   \institute{Department of Physics and Astronomy, University of Turku, 20014 Turku, Finland\\
              \email{diana.morosan@utu.fi}
            \and
       Turku Collegium for Science, Medicine and Technology, University of Turku, 20014, Turku, Finland}

  \abstract
   {Type~II radio bursts are solar radio emissions generated by electrons accelerated by coronal shocks. These bursts are typically found close to expanding coronal mass ejections (CMEs), making them valuable for studying the properties and dynamics of CME-driven shocks in the solar corona.}
   {Here, we aim to determine the regions in the solar corona where shock waves accelerate electrons and determine their characteristic properties. To do this, we combine radio observations of type~II solar radio bursts with magneto-hydrodynamic (MHD) simulations of the solar corona.}
  {We analyse ten type~II radio bursts from Solar Cycle 25 exhibiting emissions in the 150--300 MHz frequency range. The novelty of this study lies in using radio imaging data for all type~II bursts to examine the positions of the radio sources. The radio source positions, combined with a geometrical fitting of the CME shock and the MHD simulations, are used to determine essential shock parameters at the acceleration region, such as the Alfvén Mach number and shock normal angle relative to the magnetic field. The shock parameters are then combined with the properties of the radio emission and the associated eruption in a comprehensive study.}
  {We found that for all events, the type~II bursts are located near or inside coronal streamers. These streamer regions have a low Alfvén speed while the estimated shock speeds are high, resulting in the formation of supercritical shocks, with Alfvén Mach numbers in the range 3.8 to 7.7. However, in most events, type~II bursts are located at oblique shocks rather than near-perpendicular geometries, suggesting that the shock structure is more complex at local scales than the simple spherical shock models usually applied to CME shocks.}
  {Our results suggest that CME-streamer interaction regions are necessary for the generation of type~II bursts, as they provide ideal plasma conditions for the formation of supercritical shocks and the subsequent acceleration of electrons.}


   \maketitle

\section{Introduction}

    Solar type~II radio bursts can be attributed to the propagation of collisionless shocks in the solar atmosphere, often driven by energetic eruptive events, such as coronal mass ejections \citep[CMEs; ][]{nelson_type_1985, mann_catalogue_1996}. These bursts are most frequently detected in association with CMEs \citep{zucca_shock_2018, mancuso_three-dimensional_2019, morosan_extended_2020} and, in rare cases, they have also been reported in connection with shocks occurring without a CME eruption \citep[e.g.][]{morosan_type_2023}. Type~II bursts are also closely associated with extreme ultraviolet (EUV) waves, which are linked to large-scale fast magnetosonic magneto-hydrodynamic (MHD) waves and/or shocks in the solar corona \citep{klassen_catalogue_2000, warmuth_multiwavelength_2004, mann_kinematical_2023}. In the dynamic spectra, type~II bursts manifest as slowly drifting features (typically less than $1~\rm{MHz/s}$), originating at higher frequencies and gradually shifting towards lower frequencies. They are emitted at both the fundamental ($f_{\rm pe}$) and harmonic ($2f_{\rm pe}$) of the plasma frequency, with the two lanes drifting almost in parallel to each other. Type~II bursts are usually composed of many fine structures in a temporally well-resolved dynamic spectrum \citep{magdalenic_fine_2020}. Among these, the \q{herringbone} bursts appear as narrow bursts of radiation with high frequency drift rates (for example, Fig.~\ref{herringbones}). These distinct features, which exhibit either positive or negative drifts in the spectrum, are often considered signatures of individual electron beams escaping shocks \citep{holman_solar_1983, carley_low_2015, mann_radio_2018}. Herringbone bursts have been imaged only on rare occasions and show a close association with the CME or shock expansion in the corona \citep{morosan_shock-accelerated_2022, morosan_connecting_2024}. 
    
    Type~II bursts are generated when shock-accelerated non-thermal electrons undergo non-linear wave interactions, transferring their energy to Langmuir waves, which subsequently generate the observed radio emissions \citep{nelson_type_1985}. The sources of these bursts are found in the upstream regions of the shocks \citep{bhunia_imaging-spectroscopy_2023, normo_imaging_2025}. The ambient coronal conditions surrounding these sources, including the Alfvén speed ($v_{\rm A}$) and magnetic field configuration, play a crucial role in governing the intensity of the resulting radio emissions \citep{mann_characteristics_1995, kouloumvakos_coronal_2021, jebaraj_generation_2021}. Additionally, magnetic field structures such as streamer belts surrounding the source regions also influence the morphology of type~II bursts, often manifesting as changes in the drift rate observed in the dynamic spectrum 
    \citep{feng_diagnostics_2013, kong_possible_2015, koval_morphology_2023}. The vicinity of coronal streamers provides favourable conditions for type~II generation, as shocks traversing regions of reduced $v_{\rm A}$ are more likely to become supercritical \citep{mancuso_three-dimensional_2019}. Key shock parameters, including the Alfvén Mach number ($M_{\rm A}$) and the angle between the magnetic field and the shock normal ($\theta_{\rm BN}$), govern the efficiency of electron acceleration via the shock drift acceleration (SDA) mechanism \citep{mann_excitation_2022}. The electron acceleration is more efficient for near-perpendicular shocks ($\theta_{\rm BN} \sim 90^\circ$) that are supercritical. Therefore, it is important to study these parameters to comprehend the conditions under which shocks can efficiently generate observable type~II radio emissions.
    
    Only a limited number of studies have investigated the physical parameters of ambient plasma and the associated shocks using radio imaging of type~II bursts in the solar corona \citep{zucca_shock_2018, kouloumvakos_coronal_2021, morosan_connecting_2024}, while others, such as \cite{jebaraj_using_2020}, have employed radio triangulation techniques to determine the source positions and properties of type~II bursts in interplanetary space. We complement the ongoing studies by conducting a comprehensive investigation of multiple type~II events in metric wavelengths. 

    In this study, we used radio imaging observations to derive the source locations of ten type~II bursts from Solar Cycle 25 and determined the ambient plasma properties around their sources. Section \ref{Obs_data_analysis} provides an overview of the observational data and analysis methodologies employed in this study. The results are presented in Section \ref{results} and are further discussed in Section \ref{discussion}. 

\section{Observation and data analysis} \label{Obs_data_analysis}

    \begin{figure*}
    \sidecaption
        \includegraphics[width=12cm]{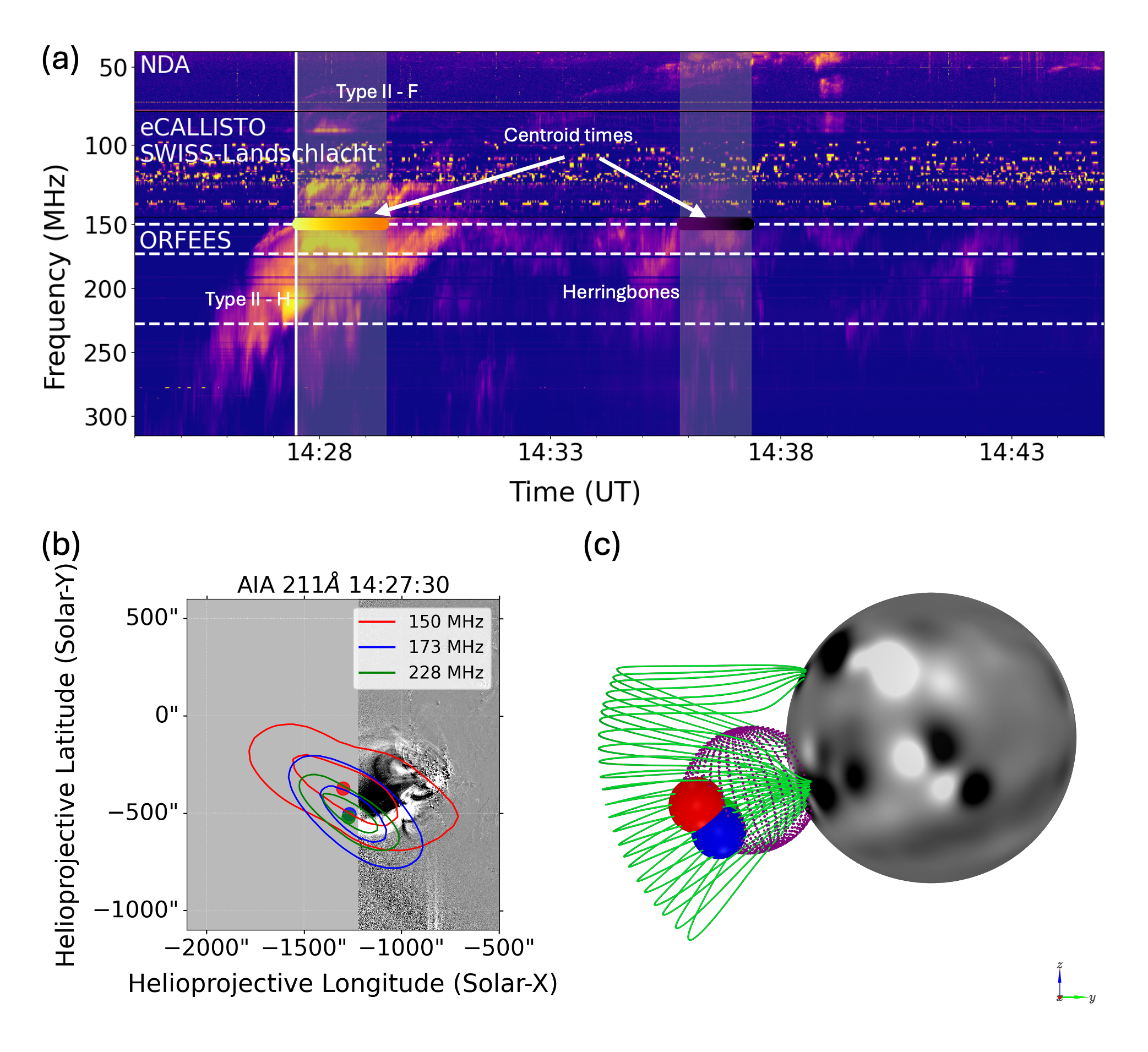}
        \caption{Composite dynamic spectrum and radio source locations from May 29, 2024. Panel a: Dynamic spectrum (315–40 MHz) combines NDA, e-CALLISTO Landschlacht, and ORFEES data. Horizontal dashed lines mark the NRH imaging frequencies used for panel b at the time indicated by the vertical dashed line. The deprojected centroids shown in Figs.~\ref{Radio_shock_sources}~and~\ref{CME_shock_normal} are overplotted on the dynamic spectrum using their respective colours to indicate their time and frequency. Panel b: Type~II radio source contours are shown on an AIA 211$\AA$ running-difference image of the associated CME eruption. Contours mark different frequencies—150 (red), 173 (blue), 228 MHz (green)—matching dashed horizontal lines in panel a, and are drawn at 40\% and 80\% of maximum intensity. Panel c: Deprojected radio source positions from panel b are plotted on the photospheric magnetogram, along with closed magnetic field lines (green) near the sources (x-axis towards Earth). The magenta mesh represents the reconstructed CME shock at 14:26 UT.}
        \label{Observations}
    \end{figure*}

    \begin{figure}[!ht]
          \resizebox{\hsize}{!}{\includegraphics{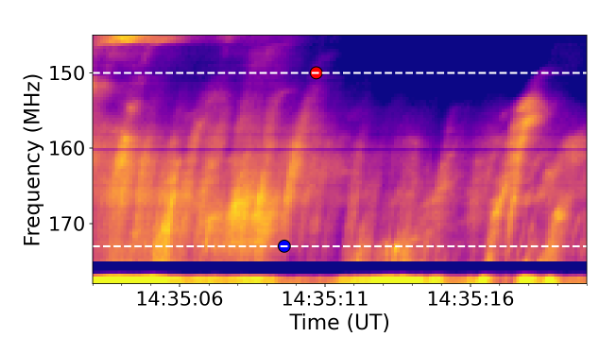}}
            \caption{Zoomed-in dynamic spectrum showing the herringbones. The blue and red dots show the start and end points of the herringbone used to estimate the electron beam speed.}
            \label{herringbones}
    \end{figure}

    \begin{table*}[htpb]
    \caption{Results of the study, including CME speeds derived from CME fitting and LASCO catalogue, shock parameters such as $\theta_{\rm BN}$ angle, Alfvén ($M_{\rm A}$) and fast magnetosonic ($M_{\rm FMS}$) Mach number, and electron beam energies obtained from the herringbone analysis.}
    
    \renewcommand{\arraystretch}{1.2} 
    \begin{tabular}{lccccccccccc}
    \hline\hline 
    \makecell{Date \\ } & 
    \makecell{Shock\\ Apex \\ speed \\ (km/s)} & 
    \makecell{Shock \\Lateral \\ speed \\ (km/s)} & 
    \makecell{LASCO \\ catalogue \\ speed \\ (km/s)} &
    \makecell{Type~II \\ duration \\ (mins)} & 
    \makecell{Type~II \\  frequency \\ range \\ (MHz)} & 
    \makecell{GOES \\ flare \\ class} & 
    \makecell{Electron \\ beam \\ energy \\ (keV)} & 
    \makecell{$\boldsymbol{\theta}_{\rm BN}$ \\ $\boldsymbol{(^\circ)}$} & 
    \makecell{Alfvén \\ Mach \\ number \\ $\boldsymbol{M}_{\boldsymbol{\rm A}}$} &
    \makecell{Fast \\ magnetosonic \\ Mach \\ number \\ $\boldsymbol{M}_{\boldsymbol{\rm FMS}}$} 
    \\
    
    \hline
     2021-06-09  & 811  & 787 & 282 & 19 & 475-20 & C1.7*  & 22  & 40-50 & 7.4, A & 3.4 \\
     2022-03-28  & 880  & 625 & 312 & 14 & 240-20 & M4.1   &  3  & 60-80 & 6.3, L & 3.1 \\
     2022-04-22  & --   & --  & --  &  6 & 290-60 & M3.4   & 16  & --    & --     & -- \\
     2022-05-19  & 1004 & 728 & 277 & 23 & 640-40 & X1.4\(^\dagger\) & 48 & 40-50 & 7.7, A & 3.9 \\
     2022-11-19  & 575  & 402 & 276 & 20 & 240-20 & M1.6   & 3   & 20-30 & 5.2, A & 2.7 \\
     2024-02-08  & --   & --  & --  & 9  & 280-40 & C9.8   & 9   & --    & --     & -- \\
     2024-03-10  & 1084 & 831 & 339 & 15 & 230-40 & M7.4   & 7   & 50-60 & 5.9, L & 3.5 \\
     2024-05-29  & 684  & 458 & 583 & 15 & 295-30 & X1.4   & 3   & 70-90 & 5.3, A &  2.9 \\
     2024-11-10  & 822  & 435 & 325 & 24 & 250-20 & M9.5   & 3   & 40-60 & 3.8, A & 2.9 \\
     2025-03-11  & --   & --  & --  & 8  & 650-50 & M1.1   & 9   & --    &  --    &  --  \\
    \hline
    \end{tabular}
    \label{results_table}
    \vspace{2mm}
    
    \captionsetup{justification=raggedright, singlelinecheck=false}
    \caption*{Note: * denotes the far-side flare event that was partially observed by GOES. \(^\dagger\) denotes the far-side flare event with the GOES flare class estimated by STIX/SoLO. Column 10: Alfvén Mach numbers; A~=~calculated using the shock apex speed; L~=~using the shock lateral speed.
    }
    \end{table*}

    We conducted a comprehensive analysis of the type~II bursts associated with coronal shock waves using data from various observatories and spacecraft. In total, ten type~II events from Solar Cycle 25 were analysed. The events were selected based on the availability of radio imaging observations from the Nançay Radioheliograph \citep[NRH; ][]{kerdraon_nancay_1997}. To minimise ionospheric effects, events from the European winter months were generally excluded. However, when winter events were considered, only those occurring near local noon were selected, ensuring that the Sun was close to its maximum altitude in the sky to minimise ionospheric propagation effects. Of the events analysed in this study, eight exhibited a clear association with a CME. The remaining two events were associated with a flare and a propagating EUV wave (April 22, 2022 and March 11, 2025).  
    
    The radio spectrum data used in this study primarily came from  Observations Radiospectrographiques pour FEDOME et l’Etude des Eruptions Solaires \citep[ORFEES; ][]{hamini_orfees_2021} at frequncies of $144 - 1004$ MHz and are supplemented at lower frequencies (less than $144~\rm MHz$) by other observatories, such as the Nançay Decameter Array \citep[NDA; ][]{lecacheux_nancay_2000}, the LOw Frequency ARray \citep[LOFAR; ][]{van_haarlem_lofar_2013, zhang_incremental_2022} and several e-CALLISTO stations \citep{benz_callisto_2005} such as Landschlacht, Germany (DLR), Alexandria, and HUMAIN. The Stokes I images from NRH at frequencies 150, 173 and 228~MHz are used to discern the location of the type~II sources. 
    
\subsection{Type~II burst on May 29, 2024} \label{example_event}

    In this subsection, we discuss the analysis methods used in the study. To do so, we use an example event on May 29, 2024 (Fig.~\ref{Observations}). The same methods are employed for the remaining events that are analysed in the study (see Appendix Figs.~\ref{09Jun21}~to~\ref{11Mar25}). 

    A type~II radio burst was seen with an accompanying halo CME on May 29, 2024, from the southeast limb of the Sun. The event was linked to an X1.4-class flare that began at 14:11 UT and peaked at 14:37 UT. The CME in this case was observed by multiple spacecraft, such as Solar Dynamics Observatory \citep[SDO; ][]{pesnell_solar_2012}, Solar and Heliospheric Observatory \citep[SOHO; ][]{domingo_soho_1995}, Solar Ultraviolet Imagers (SUVI) on board the Geostationary Operational Environmental Satellite~-$16$ \citep[GOES-16;][]{darnel_goes-r_2022}, and Solar Terrestrial Relations Observatory Ahead \citep[STEREO-A;][]{kaiser_stereo_2008}. 
    
    Fig.~\ref{Observations}a shows the composite dynamic spectrum of the type~II burst observed by ORFEES, e-CALLISTO Landschlacht, and NDA. The harmonic $(2f_{\rm pe})$ emission of the type~II burst (indicated as \q{type~II - H} in Fig.~\ref{Observations}a) is imaged by the NRH at the following frequencies: 150, 173, and 228 MHz. The radio contours are overlaid on the AIA/SDO \(211 \AA\) running-difference image in Fig.~\ref{Observations}b. A 2D Gaussian function was fitted to determine the centroid location and the peak intensity of the source. The initial type~II burst was followed by emissions lacking a distinct type~II backbone but featuring multiple herringbone bursts (indicated as \q{herringbones} in Fig.~\ref{Observations}a), exhibiting drifts towards both higher and lower frequencies. The drift rates of the herringbones were used to estimate the electron beam energy. 

    Combined EUV and white-light data enable a 3D reconstruction of the shock with near-simultaneous images from STEREO-A/EUVI and SDO/AIA, using a spherical model to achieve an optimal fit. EUV images from GOES/SUVI were also utilised, owing to the imager's larger field of view compared to AIA. This procedure was performed using PyThea \citep{kouloumvakos_pythea_2022}, an open-source software package for shock-wave reconstruction. The fitting was done at five different time steps between 14:26~UT and 14:36~UT, during which the shock wave is observed by both EUV instruments and white-light coronagraphs, SOHO/LASCO \citep{brueckner_large_1995} and STEREO-A inner-coronagraph (COR1) \citep{howard_sun_2008}. The angular separation between SDO and STEREO was roughly $15 ^\circ$. The fitting was initially done using coronagraph images, where the CME shock is more defined, providing us with constraints on the geometry. The fitted CME model is propagated backwards in time and refined to match earlier observations from the viewpoints of SDO/AIA and STEREO-A/EUVI, ensuring an optimal fit across multiple coronal heights. This approach captures the early evolution of the CME shock while preserving consistency with its later appearance in coronagraph fields of view. The fitting parameters consist of heliographic longitude and latitude in the Stonyhurst heliographic coordinate system, heliocentric distance of the spheroid centre $r_c$, radial and orthogonal axes of the spheroid, apex height, self-similar constant $\kappa$, which is the ratio of one of the semi-axes to the height of the apex from the solar surface, and $\epsilon$, which is the eccentricity of the spheroid. Table~\ref{fitting_params} contains these parameters at different times for all the events where the CME fitting was possible. Linear fits to the shock height–time and orthogonal distance–time profiles were used to estimate the apex and lateral shock speeds, considering the fits only in the interval when the shock propagated through the lower corona. From the time evolution of the fitted parameters, for this event we estimate a shock apex speed of $684~\rm km/s$ and a lateral expansion speed of $458~\rm km/s$ at the time of the type~II emission. Similar fittings were performed for the remaining events where the CME and shock outlines were visible in EUV or white-light images. Out of the eight CME events, only one event could not be fitted (February 8, 2024).

    To determine the ambient coronal conditions associated with these radio sources, we used the Magnetohydrodynamic Algorithm outside a Sphere (MAS) Thermodynamic model \citep{lionello_multispectral_2008} developed by Predictive Science Inc\footnote{\url{www.predsci.com}}. The MAS is an MHD model that uses photospheric magnetic field data, obtained from SDO/Helioseismic and Magnetic Imager \citep[HMI; ][]{scherrer_helioseismic_2012} as the boundary condition at the solar surface. The synoptic magnetograms used in the model are constructed by continuously updating magnetic field measurements along the central meridian of the Sun, thereby producing a full-Sun magnetic map. This can introduce additional uncertainties, as the magnetogram may be outdated and not representative of the magnetic field configuration at the time of the event. However, special care was taken to make sure the simulation magnetogram used for the modelling closely matched the magnetogram at the time of the observations. We used the model results to estimate coronal properties, such as electron density and magnetic field, which are then used to determine the global Alfven ($v_{\rm A}$) and fast magnetosonic ($v_{\rm FMS}$) speeds. The coronal densities were scaled by a factor of two to improve agreement with observations. This is motivated by the findings of \citet{wang_variation_2017}, who compare electron densities from the MAS model with those derived from polarised brightness (pB) observations by STEREO/COR1 \citep[see Fig.~7 in][]{wang_variation_2017}.
    
    The electron densities at various coronal heights obtained from the MAS model were used to deproject the radio centroids from the NRH images \citep[see, for example, the methods of][]{morosan_shock-accelerated_2022}. By using the electron density isosurface corresponding to the harmonic plasma frequency (\(2f_{\rm pe}\)) of $150$~MHz, the plane-of-sky radio centroids were deprojected to obtain their 3D positions, which are indicated in Fig.~\ref{Radio_shock_sources} along with the corresponding density isosurface. This density isosurface is overlaid with the Alfvén speed (\(v_{\rm A}\)) values obtained previously. Additionally, the shock geometry was assessed by deriving the angle between the magnetic field (\(\Vec{B}\)) and the shock normal (\(\Vec{\hat{n}}\)), denoted as \(\theta_{\rm BN}\). In Fig.~\ref{CME_shock_normal}, the shock geometry is outlined, where the reconstructed shock is overlaid with the values of the $\theta_{\rm BN}$ angle together with the photospheric magnetic field. Closed (green) and open (black) magnetic field lines are also shown in the figure.

    The herringbone bursts were analysed to estimate the energy of the electron beams escaping from the shock (Fig.~\ref{herringbones}). We estimated the energies by determining the speed of the electron beams from the drift rates of herringbones in the dynamic spectrum. To turn the drift rate into speed, we used the density-height profile obtained from the MAS model. For the event in Fig.~\ref{herringbones}, we get an estimate for the speed of the electron beam of \(0.10~\rm c\), which corresponds to an electron beam energy of \(2.5~\mathrm{keV}\). Similarly, the herringbone drift rate was used to estimate the electron beam energy in all the events analysed in the study and is reported in Table~\ref{results_table}. The herringbones used for this purpose are shown in the Appendix plots (Figs.~\ref{09Jun21}~to~\ref{11Mar25}).

\section{Results} \label{results}

    \begin{figure*}
    \sidecaption
        \centering
        \includegraphics[height=1\textheight]{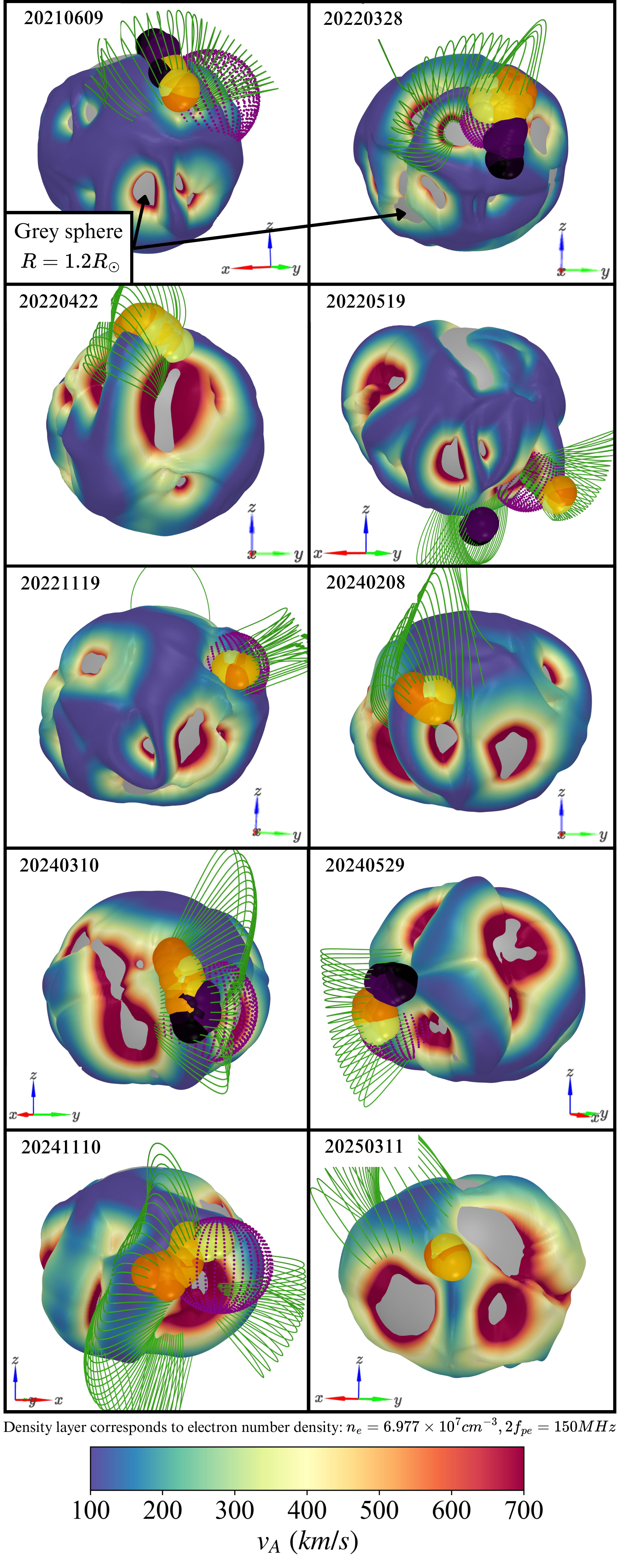}
            \caption{Density isosurface corresponding to the plasma frequency of $150$~MHz harmonic emission, overlaid with the colour map indicating Alfvén speed ($v_{\rm A}$) values. Grey sphere of radius $1.2\, R_\odot$ provides a visual reference for the height of the density isosurface. Deprojected $150$~MHz radio sources (coloured spheres in orange and purple shades) obtained from Gaussian fitting analysis (see Section \ref{Observations}), shock reconstruction (magenta mesh), and closed magnetic field lines (green) are overlaid on the density isosurface. The times and frequencies corresponding to the radio centroid positions are overlaid on the dynamic spectra shown in Figs.~\ref{Observations}~and~\ref{09Jun21}~–~\ref{11Mar25}. The coordinate axis at the bottom indicates the orientation of the Sun with the x-axis pointing towards Earth.}
            \label{Radio_shock_sources}
    \end{figure*}

    \begin{figure*}
    \sidecaption
        \includegraphics[height=1\textheight]{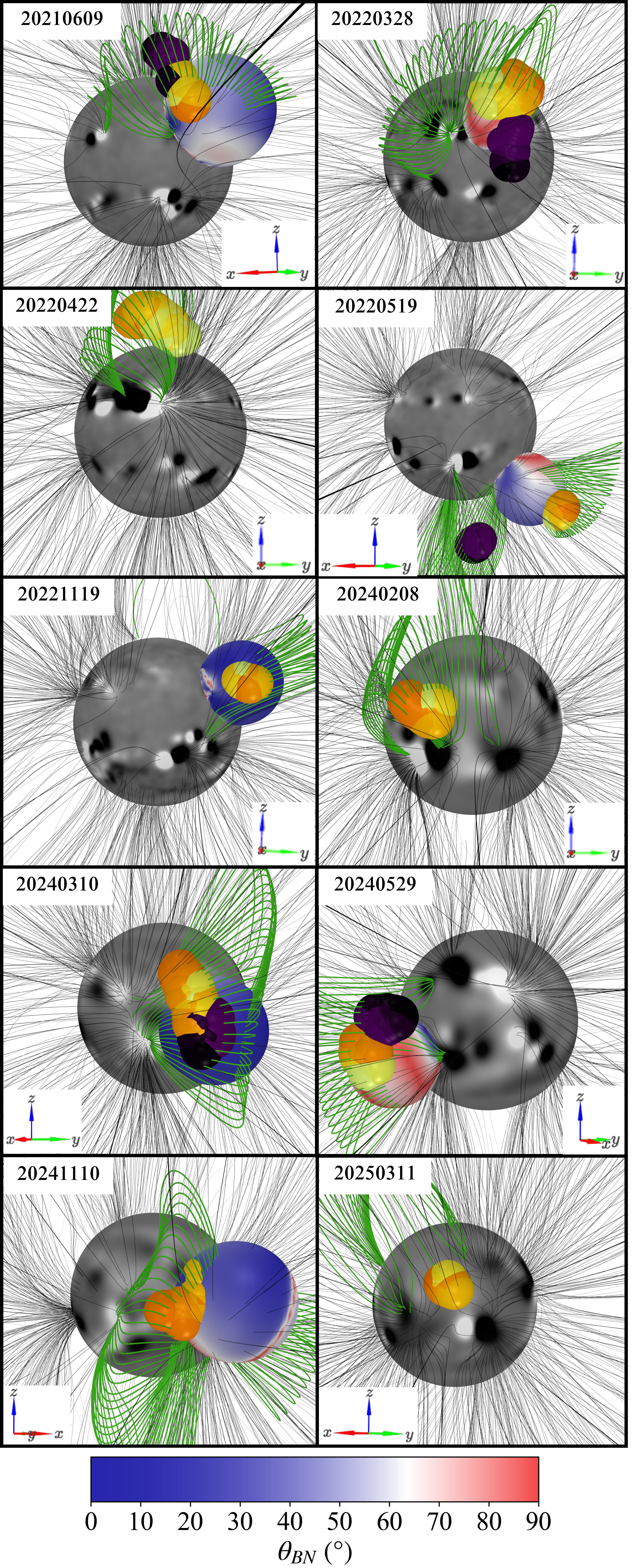}
        \caption{Shock reconstruction in 3D (when visible) for all ten events included in the study. The reconstruction is timed to coincide with the onset of the type~II radio burst or the earliest instance at which the shock reconstruction was possible. The time of the shock reconstruction for all the events is indicated as a solid vertical line in their respective spectra in Figs.~\ref{Observations}~and~\ref{09Jun21}~-~\ref{11Mar25}, and their fitting parameters have been emboldened in Table~\ref{fitting_params}. The shock reconstruction is overlaid with $\theta_{\rm BN}$ values, along with the deprojected $150$~MHz radio sources (coloured spheres in orange and purple shades) obtained from Gaussian fitting analysis (see Section \ref{Observations}). Photospheric magnetogram overlaid with both open (black) and closed (green) magnetic field lines from MAS simulation.}
        \label{CME_shock_normal}
    \end{figure*}
    
    In this section, we present our findings and common properties of the type~II bursts and those of the shock at the acceleration regions:

    To determine the type~II location relative to the shock, a spherical fitting for the shocks was performed for seven out of ten bursts, as they had an associated CME that was well-observed remotely and from multiple viewpoints. Whereas for the remaining three events, geometric fitting was not performed for the following reasons: the type~II burst on February 8, 2024, was associated with a CME moving on the disc in the North direction when observed from Earth, making it difficult to fit a spherical model. During the events on April 22, 2022, and March 11, 2025, no CME was recorded. The shocks were fitted at different times, which allowed us to evaluate their evolution and shock speed. The apex and lateral shock speeds derived from the fits are reported in Table~\ref{results_table}. We find that the shock speeds associated with these events are all fast, with apex and lateral speeds exceeding $500~\rm km/s$ and $400~\rm km/s$ respectively, in the low corona.
    
    The flare classes corresponding to the type~II bursts are reported in Table~\ref{results_table}. For events originating on the visible side of the Sun, all associated flares were relatively strong, with classes of at least C9.8. However, for far-side events (such as June 9, 2021 and May 19, 2022), the reported GOES flare classes may not accurately represent their true magnitudes. In particular, the May 19, 2022 flare, observed by the Spectrometer/Telescope for Imaging X-rays (STIX) onboard Solar Orbiter, was estimated to correspond to a GOES flare class of \(X1.9^{+X2.8}_{-M7.7}\). The event on June 9, 2021, also occurred on the far side of the Sun and was partially observed by the GOES satellite. The type~II durations are also reported in the table, and we obtain a range of $6-23$~minutes. Two events with the shortest emission times (April 22, 2022, and March 11, 2025) did not have an associated CME. The remaining events with longer durations were all associated with a CME. 

    The electron beam energies were determined by analysing the herringbones observed in each event, and the results are reported in Table~\ref{results_table}. We obtain electron beam energies in the range \(3~-~48~\rm keV\), with seven out of the ten analysed herringbones exhibiting energies less than \(10~\rm keV\) (\( 0.19~\rm c\)). The electron speeds derived in our study are consistent with previous herringbone analyses, with reported averages ranging from $0.14$~c to $0.18$~c \citep{mann_electron_2005, carley_low_2015, mann_radio_2018, morosan_connecting_2024}. 
    
    The plane-of-sky radio source positions at $150$~MHz were deprojected using the corresponding density surface at the plasma emission frequency, provided by the MAS model. Fig.~\ref{Radio_shock_sources} shows the layer of the solar corona with an electron density corresponding to the harmonic plasma emission at $2f_{\rm pe}=150$~MHz (i.e. $n_{\rm e}=6.977\times10^7~\mathrm{cm}^{-3}$), as all the type~II bursts imaged here are predominantly harmonic emissions. The density surface is overlaid with Alfvén speed values obtained from the MAS electron densities and magnetic field strength. Also depicted in these panels are the closed (green) magnetic field lines obtained from the model, extending to a height of approximately $2~\rm R_\odot$. The closed field lines in the region around the type~II radio sources are shown. The panels of Fig.~\ref{Radio_shock_sources} show that the radio sources are located in the regions with low Alfvén speed ($v_{\rm A}~\lesssim~200~\rm km/s$) and also inside or next to the closed field regions. In these regions, a shock is more likely to become supercritical when it is driven by a suitably fast CME. A supercritical shock is required for the generation of the type~II burst, which means the shock speed should exceed the Alfvén speed \citep{kouloumvakos_coronal_2021}. Based on the location of the radio sources with respect to the shock, we used either the apex or lateral speed to estimate the Alfvén Mach number by taking the ratio of shock speed to the local Alfvén speed. We obtained Alfvén Mach numbers spanning the range \(3.8~<~M_{\rm A}~<~7.7\), indicating the presence of supercritical shocks in all events. We also estimated the fast magnetosonic Mach number ($M_{\rm FMS}$) for the events and found them to be in the range \(2.7~<~M_{\rm FMS}~<~3.9\). The estimated $M_{\rm A}$ and $M_{\rm FMS}$ values for seven of the events are reported in Table~\ref{results_table}. 
    
    The magnetic field values in the solar corona, provided by the MAS model, allowed us to evaluate the magnetic field configuration around the type~II sources. In Figs.~\ref{Radio_shock_sources}~and~\ref{CME_shock_normal}, the closed magnetic field lines (green) in the vicinity of the type~II sources are shown. These closed magnetic field regions are the low corona counterpart of a streamer and are characterised by higher plasma density than their surroundings, resulting in the formation of regions with lower Alfvén speed. In our study, we observe these streamers in the coronagraphs for the events that occur close to the limb. In Figs.~\ref{Radio_shock_sources}~and~\ref{CME_shock_normal}, we note that the type~II radio sources in all the events lie within or in proximity to these streamers. This suggests that shock-streamer interactions play a key role in accelerating electrons and the subsequent generation of type~II bursts. These findings align with previous studies suggesting that streamer structures can act as efficient sites for electron acceleration \citep{kong_possible_2015, kong_electron_2016}. Our results for the May 19, 2022 event align with earlier studies, as we identify two type~II radio sources, both located near closed magnetic field lines within these low-Alfvén-speed regions \citep{vasanth_wide-band_2025, zucca_source_2025}.

    The geometry of shocks associated with the type~II bursts was identified by determining the angle between the magnetic field ($\Vec{B}$) and the shock normal ($\Vec{\hat{n}}$), denoted as $\theta_{\rm BN}$. Fig.~\ref{CME_shock_normal} consists of the photospheric magnetogram overlaid with open (black) and closed (green) magnetic field lines. To depict the \(\theta_{\rm BN}\) angles at the shock, the figure also consists of the reconstructed shock surface overlaid with the $\theta_{\rm BN}$ value. The range of values corresponding to the type~II locations is reported in Table~\ref{results_table}. The derived \(\theta_{\rm BN}\) angles consistently indicate the presence of globally oblique shocks at the location of type~II bursts, with the values being mostly less than \(80^\circ\). The presence of a near-perpendicular shock is, however, expected for the acceleration of type~II burst-generating electrons \citep{mann_characteristics_1995, mann_radio_2018}.

\section{Discussions} \label{discussion}
    A comprehensive analysis of ten type~II bursts was conducted to determine the ambient plasma parameters at their locations and the associated shock properties. These events were accompanied by strong flares and fast CMEs, except for two that did not have an accompanying CME. This combined study of multiple type~II radio bursts reveals the following main findings that are common to all of these separate events:

    \begin{itemize}
        \item The type~II sources identified in this study are all located in regions of low Alfvén speed. To quantify the strength of the associated shocks, we derived the Alfvén Mach number $M_{\rm A}$ for events where the shock fitting was possible. The critical magnetosonic Mach number ($M_{\rm FMS}^*$), which determines whether a shock is supercritical or subcritical, depends primarily on $\theta_{\rm BN}$ for \(\beta_{\rm plasma}~\ll~1\) (i.e. when the plasma pressure is much lower than the magnetic pressure). \citet{edmiston_parametric_1984} reported that the critical magnetosonic Mach number $M_{\rm FMS}^*$ depends on the shock geometry, varying from about $2.76$ in the quasi-perpendicular limit to about $1.53$ for quasi-parallel shocks. Additionally, \cite{mann_excitation_2022} find that, for plasma conditions at the $25$~MHz plasma frequency level, Langmuir waves are efficiently generated by electrons accelerated by coronal shocks with Alfvén Mach numbers in the range $1.59~<~M_{\rm A}~<~2.53$. Other observational studies, such as \cite{bemporad_identification_2011, kouloumvakos_coronal_2021}, show that the onset of type~II emission closely coincides with the transition from a subcritical to a supercritical shock. The Mach numbers determined in our study are in the range  $3.8~<~M_{\rm A}~<~7.7$ and $2.7~<~M_{\rm FMS}~<~3.9$. These results indicate that the shocks analysed here are predominantly supercritical, and therefore capable of efficiently accelerating electrons. 
        
        \item The analysis of the magnetic field configuration produced by the MAS model shows that all type~II sources identified in this study lie inside or adjacent to the closed magnetic loops that constitute the low–coronal counterparts of streamers. These streamers separate regions of opposite magnetic polarity and contain enhanced-density plasma, leading to lower Alfvén speeds where shocks can become supercritical, as discussed earlier. Previous studies \citep{mancuso_coronal_2004, cho_low_2008, magdalenic_tracking_2014, jebaraj_using_2020, jebaraj_generation_2021, kouloumvakos_coronal_2021} propose that type~II bursts are generated when a CME-driven shock propagates within or in proximity to streamer regions. Simulations of coronal shocks interacting with streamers further show that such regions can efficiently accelerate protons to energies exceeding $100$~MeV, facilitated by the quasi-perpendicular shock geometry and the trapping effects created by the closed magnetic field structures \citep{kong_acceleration_2019}. These simulation results are consistent with the observations of \cite{frassati_acceleration_2022}, who report high-energy (of approximately~$100$~MeV) particles during shock propagation through streamer-like regions, reinforcing the idea that streamers provide favourable conditions for efficient particle acceleration. Streamer-like closed magnetic fields are also known to efficiently trap and accelerate electrons by forcing multiple shock interactions during coronal shock expansion \citep{kong_possible_2015, kong_electron_2016}. Our results, therefore, align closely with previous observational and simulation studies. The localisation of type~II sources within low–Alfvén speed streamer structures, the likelihood of supercritical shock formation, and the associated favourable shock geometry are all consistent with the shock–streamer interaction scenario proposed in earlier works. 

        \item The shock geometry around the type~II sources shows that most events are associated with globally oblique shocks, with $\theta_{\rm BN}$ values typically below $\sim 80^\circ$. A near-perpendicular shock geometry is required for the acceleration of electrons via the SDA mechanism \citep{mann_radio_2018}, and consequently, the generation of type~II bursts \citep{kouloumvakos_coronal_2021}. Our findings suggest that global shock reconstructions, such as spherical or spheroidal shock fits, may not reliably capture the local geometry at the type~II source regions. An explanation for the presence of an oblique shock instead of a quasi-perpendicular shock can be found in the discussion section of \cite{morosan_connecting_2024}, where the authors consider two possible scenarios. In the first, the shock front is intrinsically non-planar, such as wavy or rippled, leading to local shock geometry alternating between quasi-parallel and quasi-perpendicular. This allows certain patches of an otherwise oblique or parallel shock to become locally quasi-perpendicular and capable of producing type~II emission. Such non-planar structures can also enable electrons to encounter the shock multiple times, gaining energy at each crossing \citep{zlobec_fine_1993, morosan_electron_2020}. Additionally, MHD simulations of shocks in interplanetary space show that the CME-driven shocks become strongly distorted as the shock propagates through the inner heliosphere due to spatial variations in the upstream solar wind, leading to a non-uniform shock surface \citep{wijsen_effect_2023}. Simulations by \cite{trotta_electron_2019} show that supercritical shocks significantly boost electron energies due to the formation of ripples at the shock front, which generate small-scale, quasi-perpendicular geometries. The second scenario in \citet{morosan_connecting_2024} involves non-ideal magnetic fields with strong, large-amplitude fluctuations. These magnetic inhomogeneities can create locally perpendicular configurations even within globally parallel shocks, and have been proposed in earlier studies to explain herringbone structures \citep{mann_electron_1995} and type~II bursts \citep{clasen_electron_1998} in quasi-parallel shocks. 
    \end{itemize}

    Our study highlights the coronal conditions common to the ten observed type~II bursts. The use of radio imaging data for the study provides accurate spatial positions for the type~II sources in the solar corona, enabling the determination of ambient plasma parameters around the shock. The shared features, such as low Alfvén speeds and locations inside or near streamers, suggest that these regions provide favourable conditions for shock formation and radio emission. Our findings also suggest that spherical models may not be suitable for shock reconstructions, as they can overlook the non-uniform structures of these shock waves.
 
    In future studies, it will be important to investigate the accelerated particles in type~II events and their solar energetic particle (SEP) counterparts in in situ measurements. The connection between these shock-accelerated electrons and those measured in situ remains poorly understood. In situ measurements connected to radio burst positions in the low corona have only been done on a few occasions \citep[e.g.][]{morosan_connecting_2024}. 

\begin{acknowledgements}
      K.S.B., D.E.M., and S.N. acknowledge the Research Council of Finland project \q{SolShocks} (grant number 354409). S.N. acknowledges the Vilho, Yrj\"o and Kalle V\"ais\" al\" a Foundation of the Finnish Academy of Science and Letters. This study has received funding from the European Union’s Horizon Europe research and innovation programme under grant agreement No.\ 101134999 (SOLER). We thank the Radio Solar Database service at LESIA \& USN (Observatoire de Paris) for making the NRH/ORFEES/NDA data available. We also thank the eCALLISTO network and the LOFAR IDOLS project for the availability of radio spectra. The authors acknowledge helpful comments and suggestions from Rami Vainio, which helped to improve the presentation of the results. 
\end{acknowledgements}

\bibliographystyle{aa}
\bibliography{references}

\begin{appendix}
    \section{Observations of the type~II events analysed for the study}
\begin{figure}
         \centering
         \caption{Date: June 9, 2021}
         \includegraphics[height=0.37\paperheight]{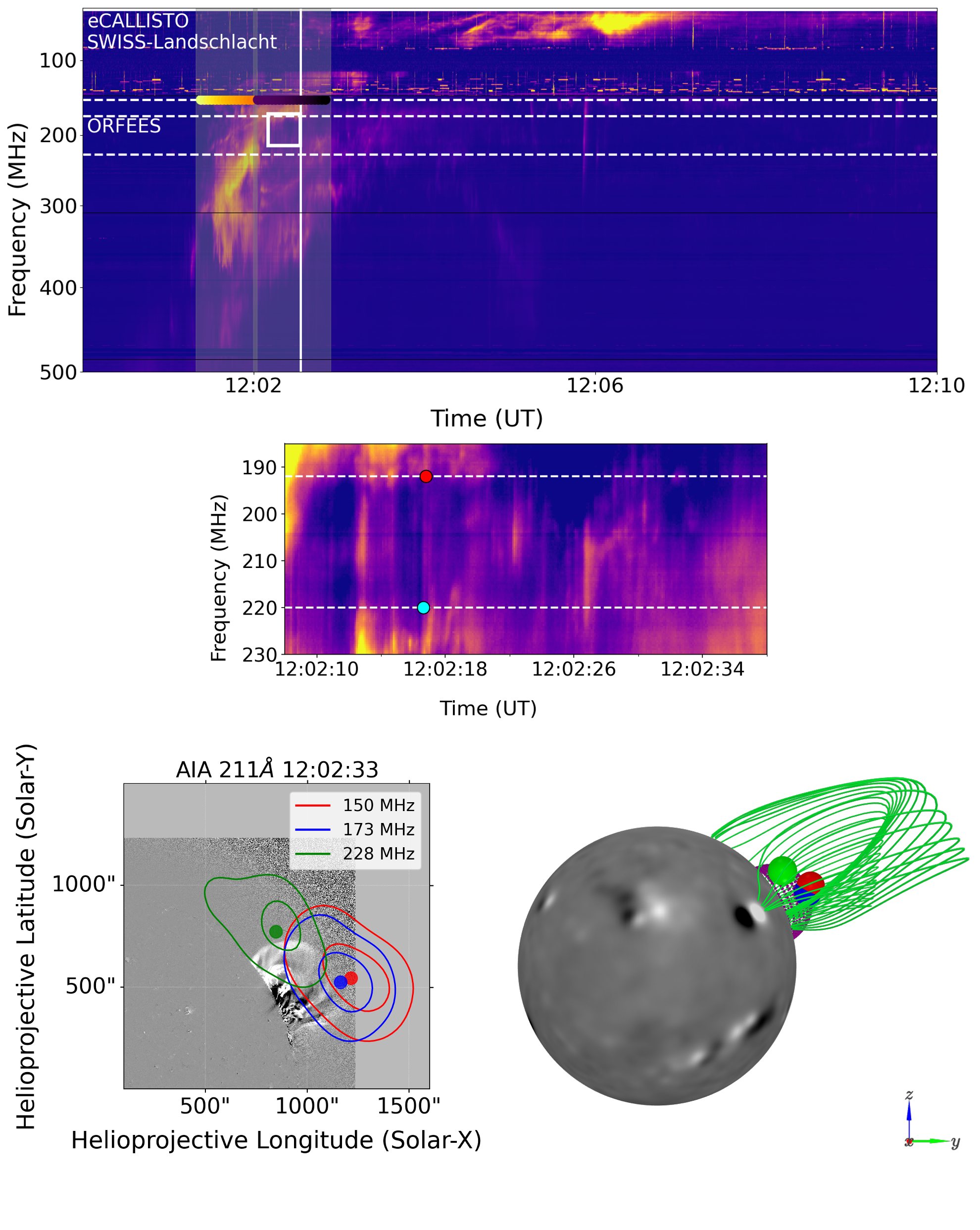}
         \label{09Jun21}
     \end{figure}

\begin{figure}
         \centering
         \caption{Date: March 28, 2022}
         \includegraphics[height=0.37\paperheight]{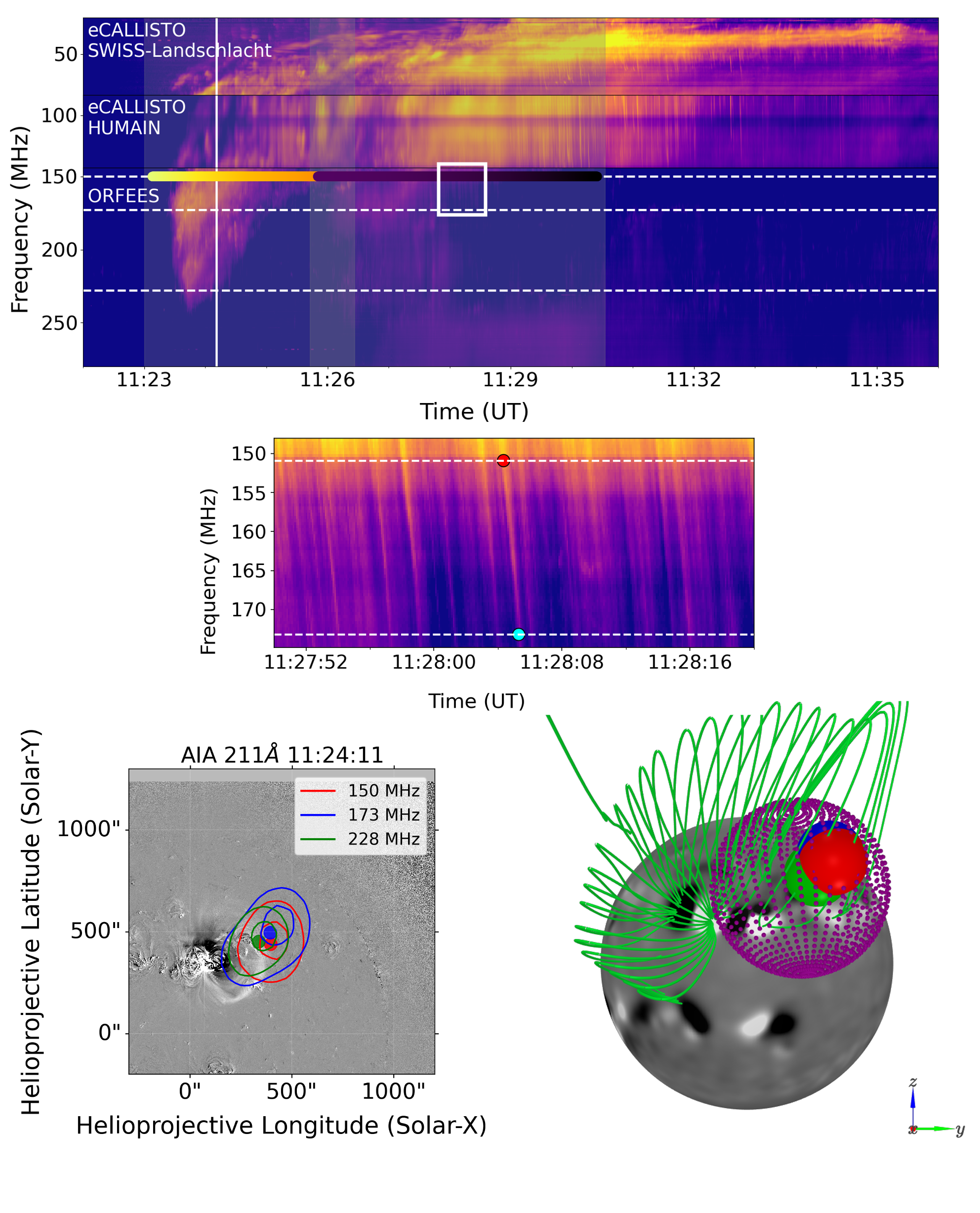}
         \label{28Mar22}
     \end{figure}

\begin{figure}
         \centering
         \caption{Date: April 22, 2022}
         \includegraphics[height=0.37\paperheight]{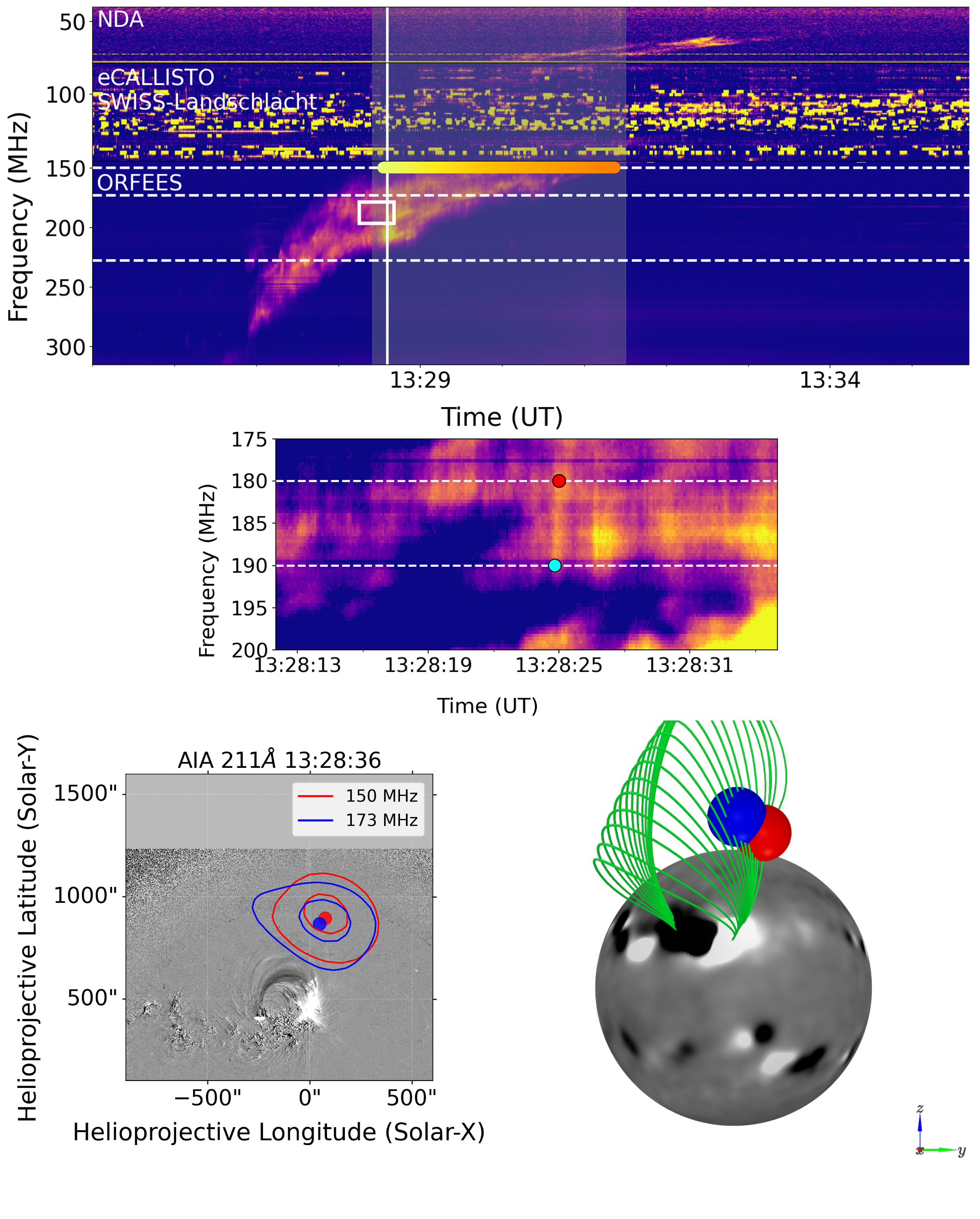}
         \label{22Apr22}
     \end{figure}

\begin{figure}
         \centering
         \caption{Date: May 19, 2022}
         \includegraphics[height=0.37\paperheight]{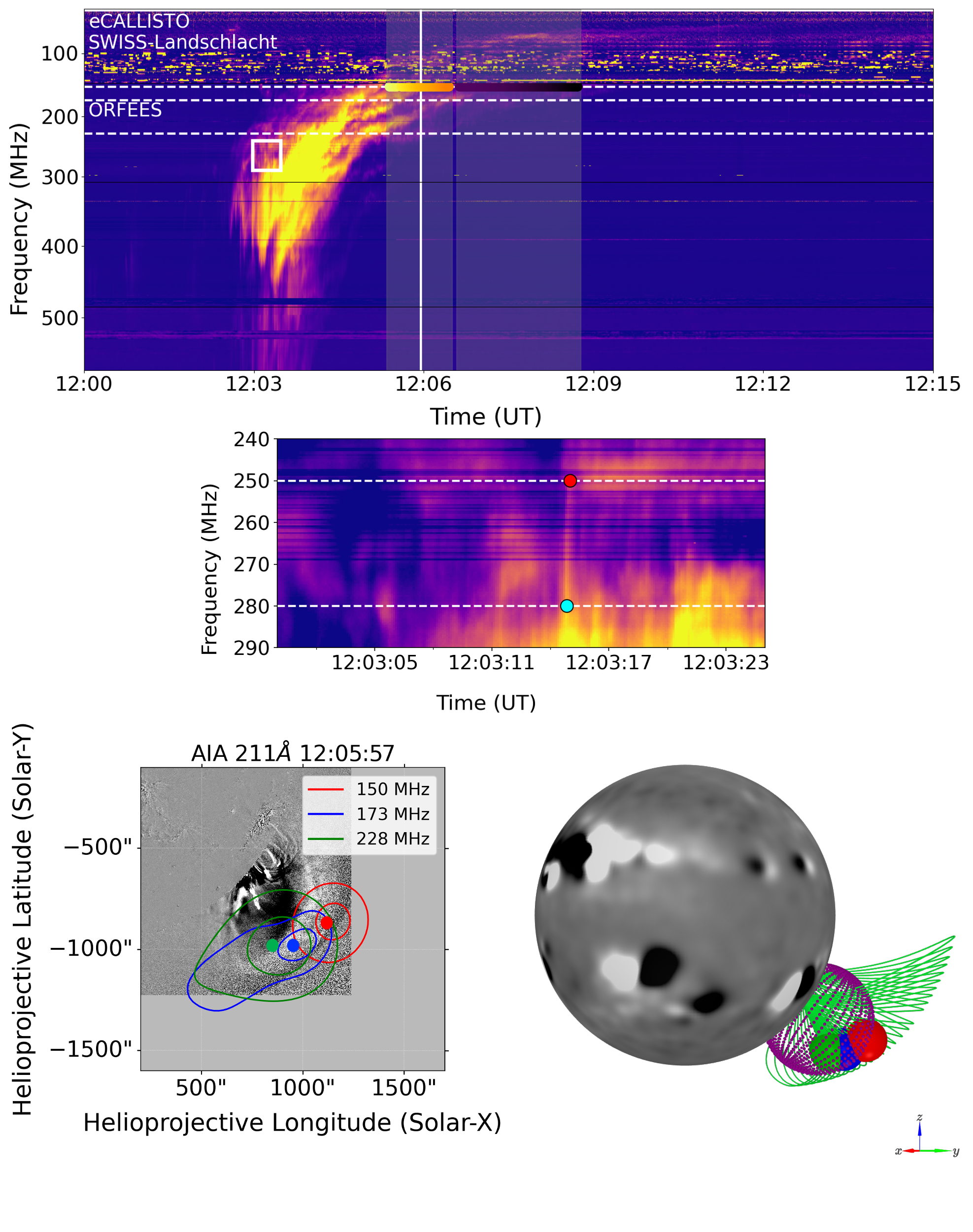}
         \label{19May22}
     \end{figure}

\begin{figure}
         \centering
         \caption{Date: November 19, 2022}
         \includegraphics[height=0.37\paperheight]{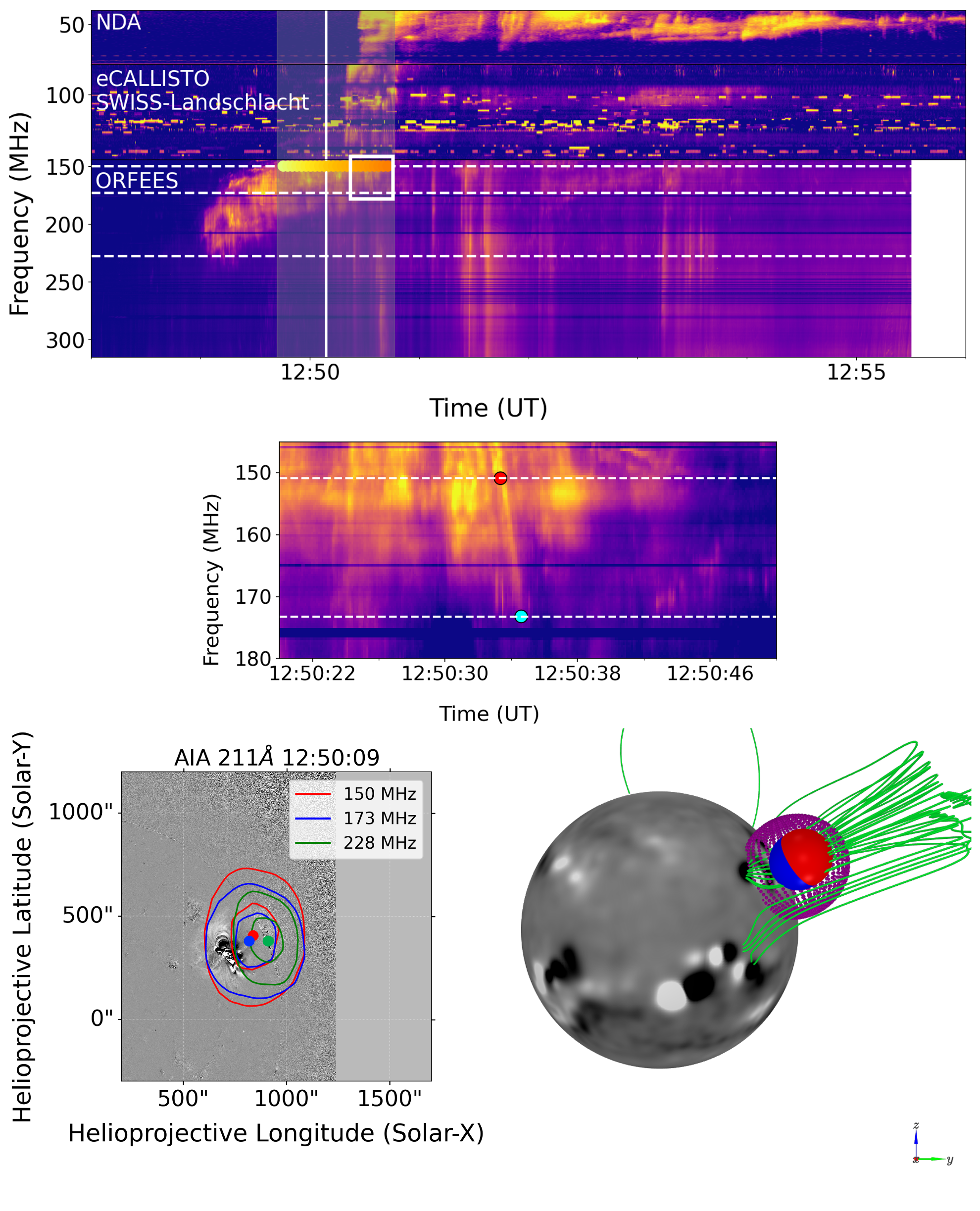}
         \label{19Nov22}
     \end{figure}

\begin{figure}
         \centering
         \caption{Date: February 8, 2024}
         \includegraphics[height=0.37\paperheight]{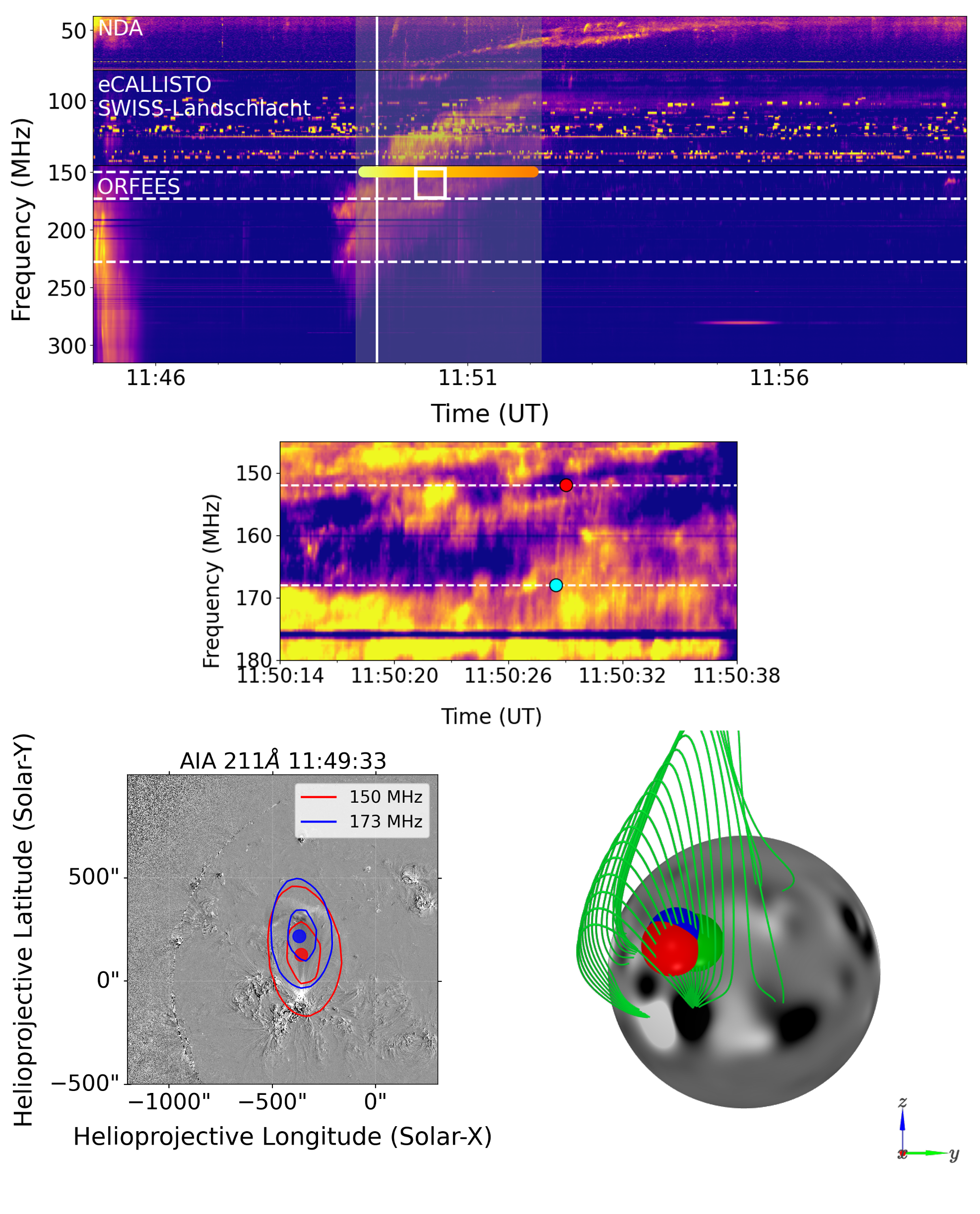}
         \label{08Feb24}
     \end{figure}

\begin{figure}
         \centering
         \caption{Date: March 10, 2024}
         \includegraphics[height=0.367\paperheight]{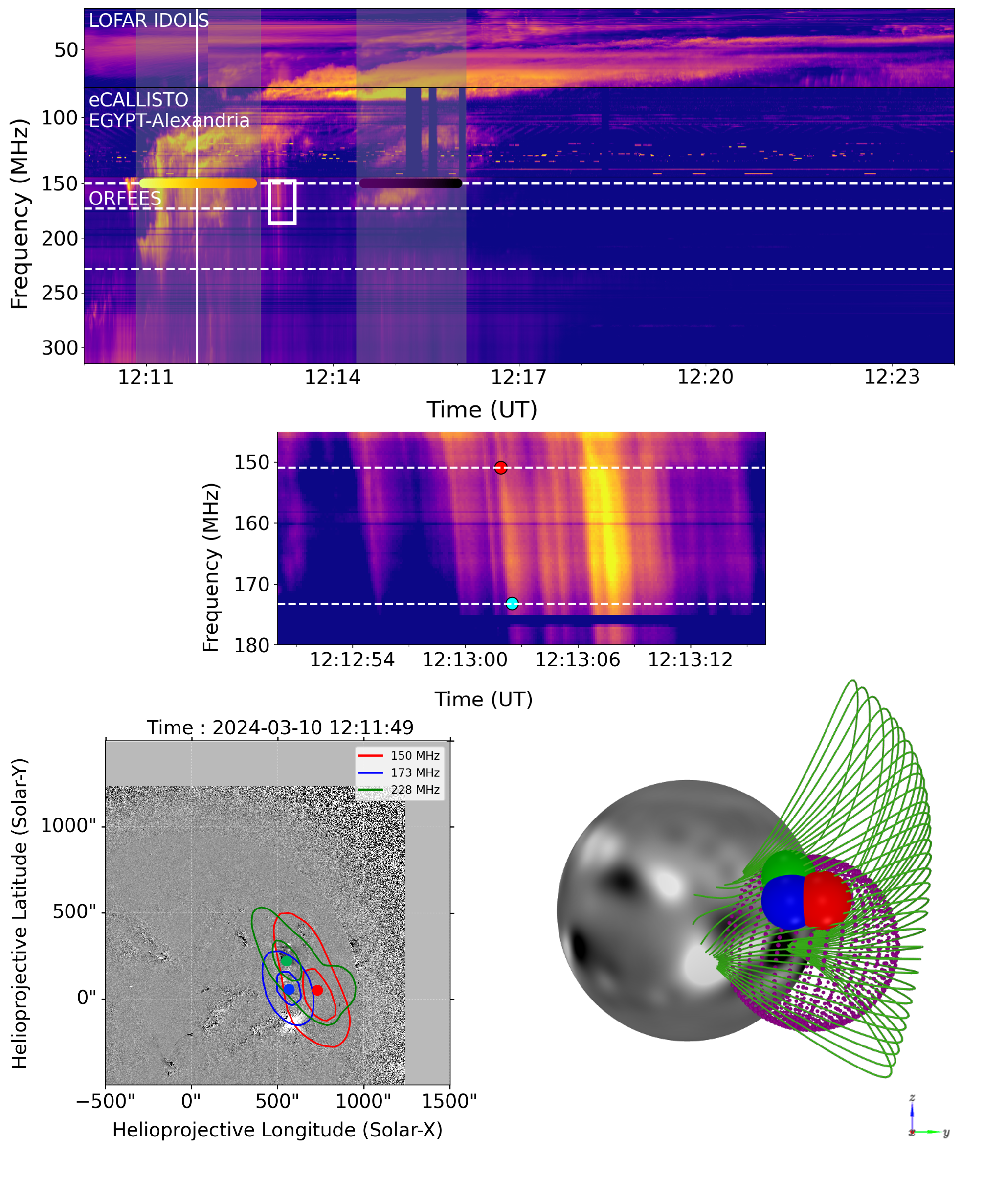}
         \label{10Mar24}
     \end{figure}

\begin{figure}
         \centering
         \caption{Date: November 10, 2024}
         \includegraphics[height=0.37\paperheight]{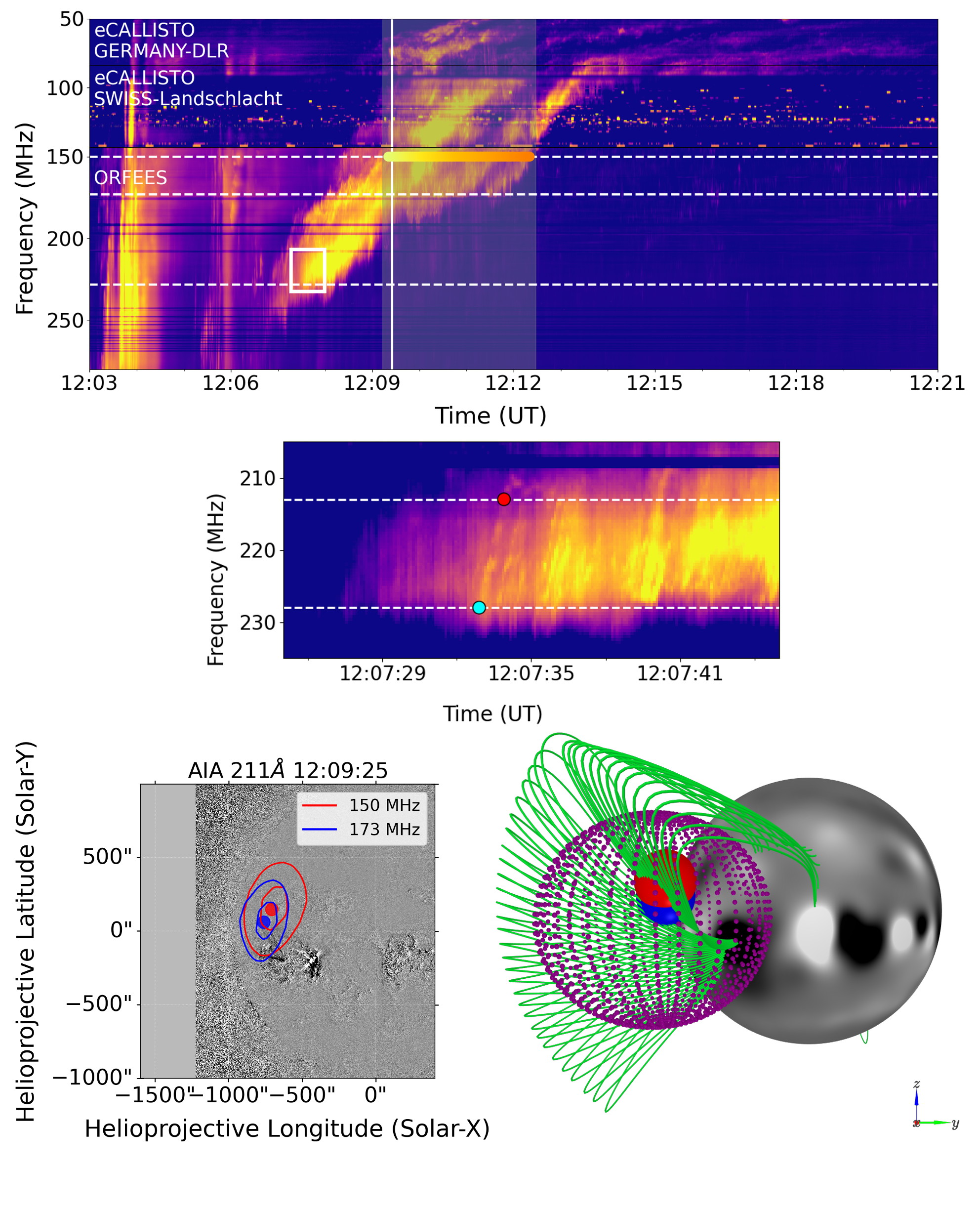}
         \label{10Nov24}
     \end{figure}

\begin{figure}
         \centering
         \caption{Date: March 11, 2025}
         \includegraphics[height=0.37\paperheight]{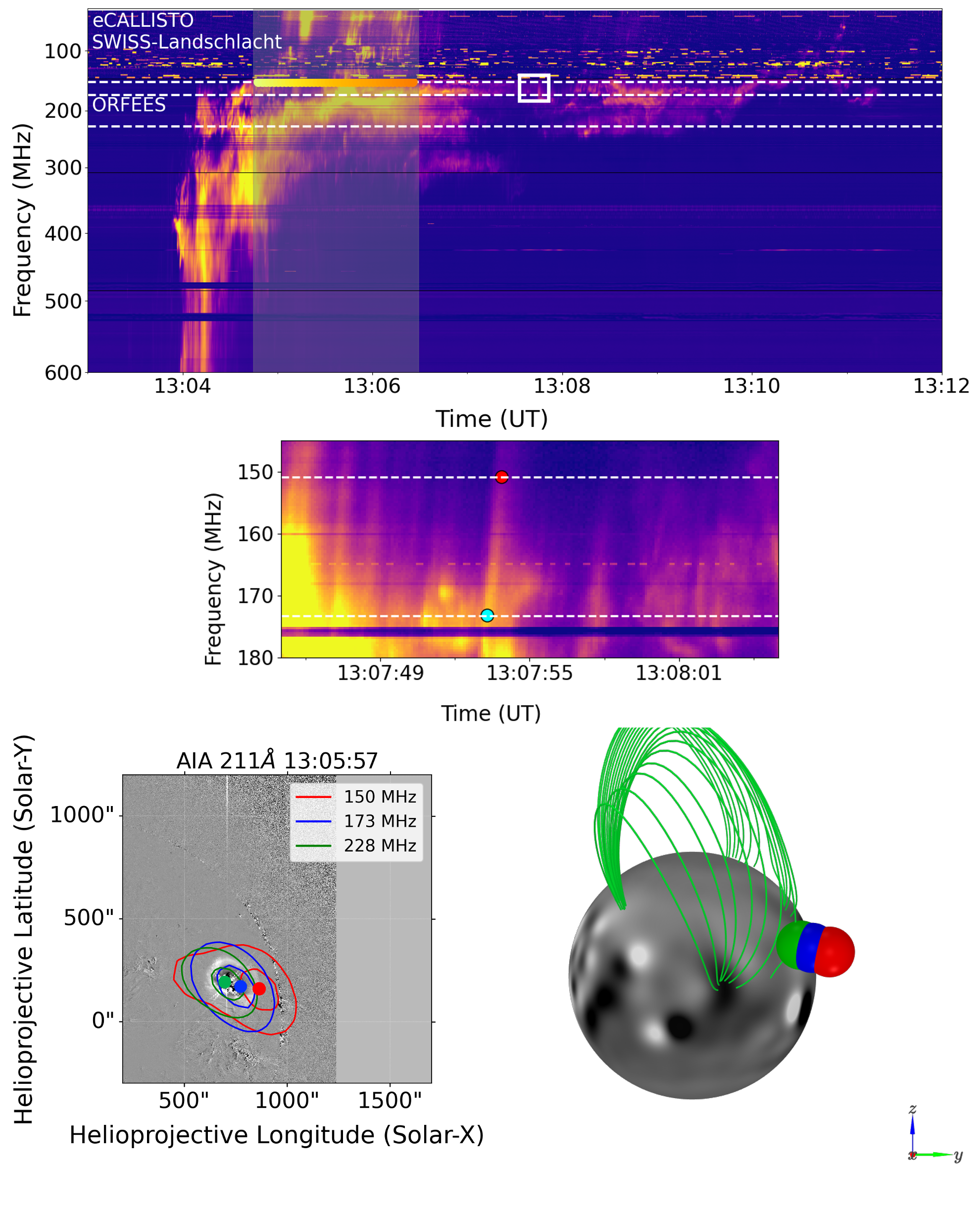}
         \label{11Mar25}
     \end{figure}

    The radio observations, including dynamic spectra and radio intensity contours at 150, 173, and 228 MHz, are presented for each event analysed in this study. Figs.~\ref{09Jun21}~to~\ref{11Mar25} show composite dynamic spectra obtained from ORFEES, NDA, LOFAR-IDOLS, and e-CALLISTO observatories. Below the dynamic spectrum is a zoom-in on the herringbone, which is used to estimate the electron beam energy. The bottom panel includes the radio image contours overlaid on the running-difference SDO/AIA $211~\AA$ images (bottom left). The time corresponding to the SDO/AIA image is marked on the dynamic spectrum with a white vertical line, while the dashed horizontal lines indicate the NRH imaging frequencies used. The fourth figure (bottom right) represents the solar magnetogram at the time of the event, which serves as the boundary condition for the MAS model. This figure also includes closed magnetic field lines (green) extending up to a height of $\sim 2R_\odot$, while the magenta mesh represents the reconstructed shock. The centroids derived from the radio images are deprojected using the density isosurface from the MAS model and plotted on this diagram. 

    \vspace{1cm}
    \noindent Caption for Figs.~\ref{09Jun21}~-~\ref{11Mar25}: Top: composite dynamic spectrum of the type~II burst. Horizontal dashed lines denote the imaging frequencies (i.e. 150, 173 and 228 MHz). The deprojected centroids shown in Figs.~\ref{Radio_shock_sources}~and~\ref{CME_shock_normal} are overplotted on the dynamic spectrum using their respective colours to indicate their time and frequency. The white box denotes the spectral region of the herringbones. Centre: zoomed-in spectral profile of the herringbone within the white box in the dynamic spectrum. Horizontal dashed line, along with the red and cyan dots, indicates the frequencies and times of the herringbones. Bottom left: SDO/AIA running-difference images at \(211\AA\) overlaid with $40\%$ and $80\%$ contours of NRH images along with the respective centroids, at the frequencies marked in the spectrum. Bottom right: deprojected radio sources from the bottom left panel, plotted on the photospheric magnetogram, along with the closed magnetic field lines (green) near the sources (as viewed from the Earth, x-axis towards Earth). The reconstructed CME shock (magenta mesh) is at the closest available time to the type~II onset at the above frequencies.
    
    \FloatBarrier
    
    \section{Shock reconstruction} 
    The shock observations from multiple spacecraft allow us to reconstruct the shock surface using a simple self-similar expanding spheroid. The fitting parameters for the spheroid surface are reported in Table~\ref{fitting_params}. The table consists of heliographic coordinates of the sphere on the solar surface, heliocentric distance ($r_c$) of the spheroid centre, along with other parameters. The table also consists of the self-similar constant ($\kappa$) and eccentricity ($\epsilon$) to describe the geometry of the spheroid. The definitions of these fitting parameters can be found in \cite{kouloumvakos_pythea_2022}. 

    \begin{table*}[!htpb]
    \caption{Spheroid fitting parameters determined using the PyThea tool for the shocks associated with the type~II bursts. Boldface entries indicate the shock fits used in Figs.~\ref{Radio_shock_sources}~and~\ref{CME_shock_normal}.}
    \centering
    \renewcommand{\arraystretch}{1.25} 
    \resizebox{2.04\columnwidth}{!}{
    \begin{tabular}{cccccccccc}
    \hline\hline
    \makecell{Date} &
    \makecell{Time \\ (UT)} & 
    \makecell{Heliographic \\ longitude} & 
    \makecell{Heliographic \\ latitude } & 
    \makecell{Heliocentric \\ distance \\ $r_c$ ($R_\odot$)} & 
    \makecell{Radial \\ axis\\ ($R_\odot$)} & 
    \makecell{Orthogonal \\ axis\\ ($R_\odot$)} & 
    \makecell{Apex \\ height\\ ($R_\odot$)} & 
    \makecell{Self \\ similar \\ constant \\ $\kappa$} & 
    \makecell{Eccentricity \\ $\epsilon$} \\
    \hline
    \multirow{7}{5em}{2021-06-09} & 12:03 & 112.34 & 26.17 & 1.13 & 0.35 & 0.34 & 1.48 & 0.70 & 0.26 \\
                                  & 12:04 & 112.34 & 26.17 & 1.15 & 0.39 & 0.38 & 1.54 & 0.70 & 0.26 \\
                                  & 12:05 & 112.34 & 26.17 & 1.12 & 0.49 & 0.48 & 1.61 & 0.78 & 0.26 \\
                                  & \textbf{12:06} & \textbf{112.34} & \textbf{26.17} & \textbf{1.13} & \textbf{0.57} & \textbf{0.55} & \textbf{1.70} & \textbf{0.78} & \textbf{0.26} \\
                                  & 12:24 & 112.34 & 26.17 & 1.52 & 1.24 & 1.20 & 2.76 & 0.68 & 0.26 \\
                                  & 12:36 & 112.34 & 26.17 & 1.67 & 1.60 & 1.54 & 3.27 & 0.68 & 0.26 \\
                                  & 12:48 & 112.34 & 26.17 & 1.83 & 1.98 & 1.91 & 3.81 & 0.68 & 0.26 \\
    \hline

    \multirow{6}{5em}{2022-03-28} & \textbf{11:25} & \textbf{13.08} & \textbf{17.75} & \textbf{1.16} & \textbf{0.31} & \textbf{0.33} & \textbf{1.47} & \textbf{0.71} & \textbf{-0.39} \\
                                  & 11:27 & 13.08 & 17.75 & 1.21 & 0.41 & 0.44 & 1.62 & 0.71 & -0.39 \\
                                  & 11:30 & 13.08 & 17.75 & 1.28 & 0.53 & 0.58 & 1.81 & 0.71 & -0.39 \\
                                  & 11:32 & 13.08 & 17.75 & 1.36 & 0.67 & 0.73 & 2.03 & 0.71 & -0.39 \\
                                  & 11:37 & 13.08 & 17.75 & 1.48 & 0.92 & 0.99 & 2.40 & 0.71 & -0.39 \\
                                  & 12:23 & 13.08 & 17.75 & 2.72 & 3.26 & 3.54 & 5.98 & 0.71 & -0.39 \\
    \hline

    \multirow{6}{5em}{2022-05-19} & 12:02 & 120.21 & -37.68 & 1.12 & 0.19 & 0.21 & 1.31 & 0.68 & -0.38 \\
                                  & \textbf{12:06} & \textbf{120.21} & \textbf{-37.68} & \textbf{1.22} & \textbf{0.43} & \textbf{0.46} & \textbf{1.65} & \textbf{0.71} & \textbf{-0.38} \\
                                  & 12:08 & 120.21 & -37.68 & 1.26 & 0.49 & 0.53 & 1.75 & 0.71 & -0.38 \\
                                  & 12:24 & 120.21 & -37.68 & 1.54 & 1.02 & 1.11 & 2.56 & 0.71 & -0.38 \\
                                  & 12:36 & 120.21 & -37.68 & 1.67 & 1.29 & 1.39 & 2.96 & 0.71 & -0.38 \\
                                  & 12:48 & 120.21 & -37.68 & 1.80 & 1.54 & 1.66 & 3.34 & 0.71 & -0.38 \\
    \hline

    \multirow{7}{5em}{2022-11-19} & 12:51 & 51.70 & 20.87 & 1.08 & 0.29 & 0.29 & 1.37 & 0.78 & 0.27 \\
                                  & \textbf{12:53} & \textbf{51.70} & \textbf{20.24} & \textbf{1.12} & \textbf{0.35} & \textbf{0.33} & \textbf{1.47} & \textbf{0.71} & \textbf{0.27} \\
                                  & 13:00 & 51.70 & 20.24 & 1.19 & 0.53 & 0.51 & 1.72 & 0.71 & 0.27 \\
                                  & 13:05 & 51.70 & 20.24 & 1.25 & 0.72 & 0.69 & 1.97 & 0.71 & 0.27 \\
                                  & 13:11 & 51.70 & 21.49 & 1.37 & 1.03 & 0.99 & 2.40 & 0.71 & 0.27 \\
                                  & 13:21 & 51.70 & 21.49 & 1.45 & 1.26 & 1.21 & 2.71 & 0.71 & 0.27 \\
                                  & 13:38 & 51.70 & 21.49 & 1.71 & 2.00 & 1.92 & 3.71 & 0.71 & 0.27 \\
    \hline

    \multirow{7}{5em}{2024-03-10} & 12:11 & 43.92 & -4.05 & 1.13 & 0.40 & 0.39 & 1.53 & 0.73 & 0.24 \\
                                  & 12:12 & 43.92 & -4.05 & 1.16 & 0.50 & 0.48 & 1.66 & 0.73 & 0.24 \\
                                  & \textbf{12:13} & \textbf{43.92} & \textbf{-4.05} & \textbf{1.17} & \textbf{0.52} & \textbf{0.50} & \textbf{1.69} & \textbf{0.73} & \textbf{0.24} \\
                                  & 12:15 & 43.92 & -4.05 & 1.20 & 0.59 & 0.58 & 1.79 & 0.73 & 0.24 \\
                                  & 12:17 & 43.92 & -4.05 & 1.29 & 0.71 & 0.69 & 1.95 & 0.73 & 0.24 \\
                                  & 12:31 & 44.68 & 1.28  & 1.48 & 1.16 & 1.13 & 2.64 & 0.69 & 0.22 \\
                                  & 12:36 & 44.68 & 1.28  & 1.54 & 1.30 & 1.27 & 2.84 & 0.69 & 0.22 \\
                                  & 12:41 & 44.68 & 1.28  & 1.60 & 1.44 & 1.41 & 3.04 & 0.69 & 0.22 \\
    \hline

    \multirow{5}{5em}{2024-05-29} & \textbf{14:26} & \textbf{-69.62} & \textbf{-16.62} & \textbf{1.20} & \textbf{0.40} & \textbf{0.40} & \textbf{1.60} & \textbf{0.67} & \textbf{-0.18} \\
                                  & 14:29 & -69.62 & -16.62 & 1.26 & 0.49 & 0.50 & 1.75 & 0.67 & -0.18 \\
                                  & 14:31 & -69.62 & -16.62 & 1.29 & 0.56 & 0.57 & 1.85 & 0.67 & -0.18 \\
                                  & 14:33 & -69.62 & -16.62 & 1.34 & 0.66 & 0.67 & 2.00 & 0.67 & -0.18 \\
                                  & 14:36 & -69.62 & -16.62 & 1.43 & 0.82 & 0.84 & 2.25 & 0.67 & -0.18 \\
    \hline
    
    \multirow{6}{5em}{2024-11-10} & \textbf{12:10} & \textbf{-33.24} & \textbf{-2.20} & \textbf{1.35} & \textbf{0.40} & \textbf{0.40} & \textbf{1.75} & \textbf{0.53} & \textbf{0.14} \\
                                  & 12:12 & -33.24 & -2.20 & 1.44 & 0.50 & 0.50 & 1.94 & 0.53 & 0.14 \\
                                  & 12:16 & -33.24 & -2.20 & 1.55 & 0.64 & 0.63 & 2.19 & 0.53 & 0.14 \\
                                  & 12:21 & -33.24 & -2.20 & 1.68 & 0.79 & 0.78 & 2.47 & 0.53 & 0.14 \\
                                  & 12:26 & -33.24 & -2.20 & 1.80 & 0.92 & 0.91 & 2.72 & 0.53 & 0.14 \\
    \hline
    \end{tabular}
    }
    \label{fitting_params}
    \end{table*}

\end{appendix}

\end{document}